\documentclass[prl,twocolumn, aps, amsmath, amssymb,superscriptaddress, footinbib,]{revtex4-1}
\usepackage{graphicx,amsmath,amssymb}
\usepackage[usenames]{color}

\begin{document}

\title{Entanglement entropy of the $\nu=1/2$ composite fermion non-Fermi liquid state.}
\author{Junping Shao}
\altaffiliation{Department of Physics, Cornell University, Ithaca, NY 14853, USA}
\affiliation{Department of Physics, Binghamton University, Binghamton, NY 13902, USA}
\author{Eun-Ah Kim}
\affiliation{Department of Physics, Cornell University, Ithaca, NY 14853, USA}
\author{F.D.M. Haldane}
\affiliation{Department of Physics, Princeton University, Princeton, NJ 08544, USA}
\author{Edward H. Rezayi}
\affiliation{Department of Physics, California State University Los Angeles, Los Angeles, CA 90032, USA}

\begin{abstract}
The so-called ``non-Fermi liquid'' behavior  is very common in strongly correlated systems. However, its operational definition in terms of ``what it is not'' is a major obstacle against theoretical understanding of this fascinating correlated state. 
Recently there has been much interest in entanglement entropy as a theoretical tool to study non-Fermi liquids. So far explicit calculations have been limited to models without direct experimental realizations. Here we focus on a two dimensional electron fluid under magnetic field and filling fraction $\nu=1/2$, which is believed to be a non-Fermi liquid state. 
Using the composite fermion (CF) wave-function which captures the  $\nu=1/2$ state very accurately, 
we compute the second R\'enyi entropy using variational Monte-Carlo technique and an 
efficient parallel algorithm. We find the entanglement entropy scales as $L\log L$ with 
the length of the boundary $L$ as it does for free fermions, albeit  with a pre-factor twice 
that of the free fermion. We contrast the results against theoretical conjectures and discuss the implications of the results.

\end{abstract}

\maketitle
Despite its ubiquity in strongly correlated materials, the metallic `non-Fermi liquid' behavior has been challenging to characterize theoretically. At the phenomenological level, non-Fermi liquid behavior is defined by a metallic system exhibiting physical properties that are qualitatively inconsistent with Landau's Fermi-liquid theory. Examples of non-Fermi liquid metals include the strange-metal phase of the high T$_c$ cuprates\cite{Boebinger30012009}, systems near a metallic quantum critical point\cite{Gegenwart:2008fk,RevModPhys.73.797,Bruin15022013} and two-dimensional electron system subject to a magnetic field at filling $\nu=1/2$ (often referred to as Fermi-liquid-like state) 
\cite{PhysRevB.40.12013,PhysRevLett.65.112,PhysRevLett.71.3846}. However, there are many ways in which a system can deviate from a normal Fermi-liquid, such as diverging effective mass, vanishing quasiparticle weight, and anomalous transport\cite{PhysRevB.47.7312,RevModPhys.73.797,PhysRevLett.63.1996,PhysRevB.78.035103,Faulkner27082010,PhysRevLett.105.151602} and little is known about how different forms of deviation can be related. Hence
the theoretical challenge of addressing a problem without a weakly interacting quasiparticle description
has been compounded by the lack of a measure that can be used to define and classify non-Fermi liquids.

Here we turn to a quantum information measure that is sensitive to entangled nature of many-body wave-functions: the bi-partite entanglement entropy.  For gapped systems, the entanglement entropy of the reduced density matrix 
 $\rho_{A}\equiv\mathrm{Tr}_{B}|\Psi\rangle \langle \Psi|$
of a subsystem $A$ with respect to its complement $B$ for a given ground state wave-function $ \left|\Psi\right\rangle$ is widely believed to follow the area law, i.e.,  asymptotically proportional to the contact
area of two subsystems, with rigorous arguments for lattice systems~\cite{PhysRevLett.100.070502,RevModPhys.82.277}. 
On the other hand, an explicit formula for a multiplicative logarithmic correction to the area law was suggested by \textcite{PhysRevLett.96.100503} based on the Widom conjecture\cite{widom} and numerically confirmed in Ref.~\cite{calabrese} for free fermions at dimensions $d>1$. This dramatic violation of the area law for free fermions is in stark contrast to the area law found for critical bosons\cite{RevModPhys.82.277}
 up to subleading corrections\cite{PhysRevB.80.115122,PhysRevB.79.115421}
and shows how Fermionic statistics by itself drives long-range entanglement in the presence of a Fermi surface.

A key question is whether non-perturbative strong correlation effects would further enhance bi-partite entanglement entropy.
Since the explicit form of bi-partite entanglement entropy found in Refs.~\cite{PhysRevLett.96.100503,calabrese} follows from exact results on non-interacting one-dimensional Fermion systems associated each points in Fermi surface\cite{PhysRevLett.105.050502}, strong interactions are likely to cause corrections to this explicit form. So far the only explicit results available for strongly interacting fermions at $d>1$ is by \textcite{vishwanath}  for Gutzwiller projected two-dimensional (2D) fermi-surface which is a candidate ground state wave-function for a critical spin-liquid with spinon Fermi surface.  Their variational Monte Carlo calculation of second R\'enyi entropy  $S_2$ showed little change in both the functional dependence on $L_A$ the linear dimension of the subsystem $A$ (i.e., $S_2\propto L_A\log L_A$) and the coefficient upon projection. Following this numerical work, \textcite{swinglesenthil} argued that 
the entanglement entropy of certain non-Fermi liquid states would be  
given by the free fermion formula of Ref.~\cite{PhysRevLett.96.100503}.

In this letter we focus on the composite fermion wave-function for the half-filled Landau level $\nu=1/2$ state proposed by \textcite{PhysRevLett.72.900} as a test case of non-Fermi liquids. Though it is a candidate wave-function, 
it is an electronic wave-function with strong numerical, theoretical and experimental support as a ground state wave-function capturing the observed $\nu=1/2$ non-Fermi liquid state. First,
the structure factor calculated with this wave-function shows a good agreement with the structure factor obtained from the exact ground state wave-function~\cite{PhysRevLett.72.900}. Further, the wave-function is supported by 
field theoretical studies of fermions coupled to Chern-Simons gauge theory~\cite{PhysRevB.47.7312,Nayak1994359}as it describes a state with diverging effective mass for fermions with flux attachment. Finally, the  $\nu=1/2$ state is experimentally established to be a non-Fermi liquid state with a Fermi surface supporting anomalous sound propagation~\cite{PhysRevB.40.12013,PhysRevLett.65.112,PhysRevLett.71.3846}, in agreement with expectations of Refs.~\cite{PhysRevB.47.7312,Nayak1994359}. We calculate second 
R\'enyi entropy $S_2$ using variational Monte Carlo techniques implementing an algorithm improved from those used in Refs.\cite{vishwanath,mcminis}. We will focus on the comparison between subsystem linear dimension $L_A$ dependence of $S_2$ for free fermions and for the CF wave-function. 
 
\begin{figure}[h]\centering
        \includegraphics[width=.25\textwidth]{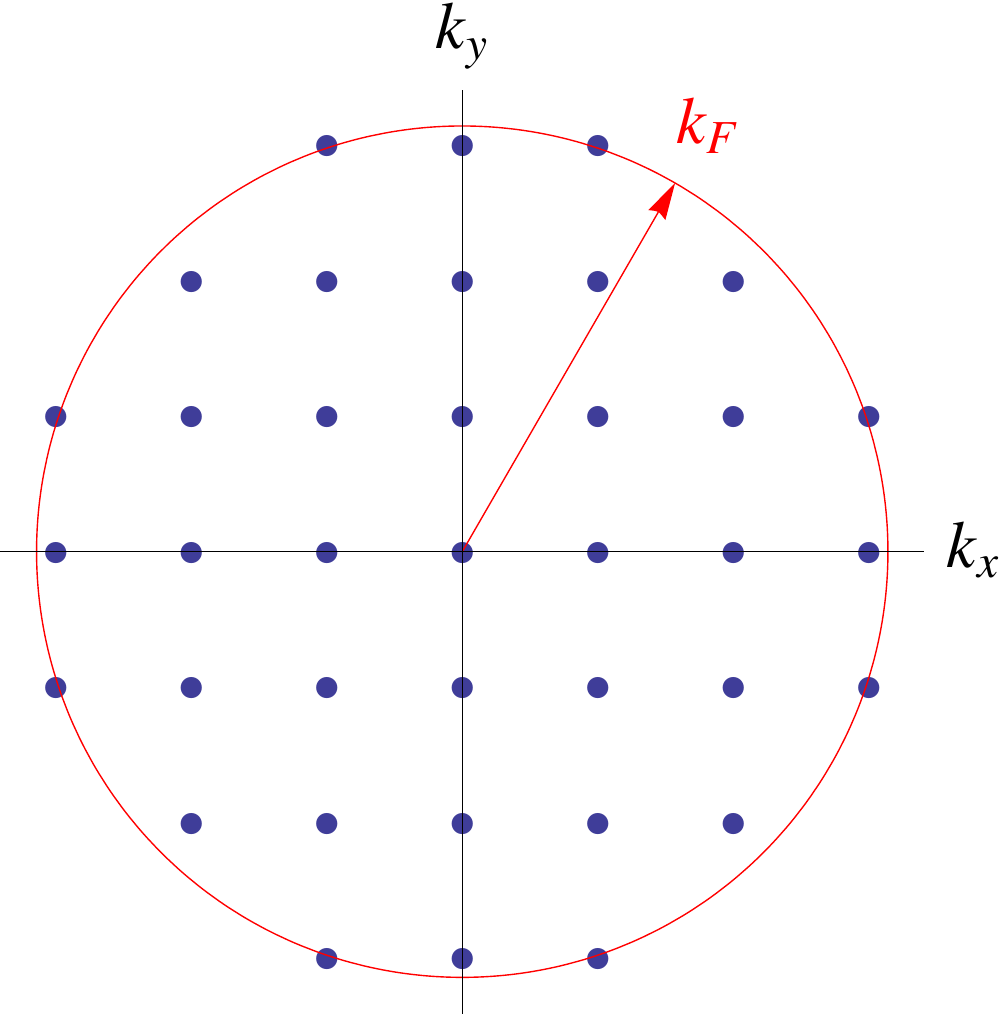}
        \caption{{Fermi surface for $N=37$ particles in 2D.} The set of momenta are shown as blue points. Red circle denotes the Fermi surface $\partial\Gamma$ of radius {$k_F \approx \sqrt{10\pi/37}$}.  
          }
        \label{fig:fermisurface}
      \end{figure}

{\it R\'enyi entanglement entropy and Widom formula.--} The second R\'enyi entropy  
is defined as
 \begin{equation}
 S_{2}\equiv -\ln\left[\mathrm{Tr}_{A}\left\{ \rho_{A}^{2}\right\} \right],
 \end{equation}
 where  
$\rho_{A}\equiv\mathrm{Tr}_{B}|\Psi\rangle\langle \Psi|$
is the reduced density matrix of region $A$. 
$S_2$ has become a quantity of growing interest as a measure of bi-partite entanglement since a convenient scheme for calculating $S_2$ using variational Monte Carlo technique was shown in Ref.~\cite{melko2010}.  
For free fermions the leading $L_A$ dependence of second R\'enyi entropy is given by\cite{PhysRevLett.96.100503,calabrese} 
\begin{align}\label{eq:klich}
 S_2&=\dfrac{3}{4}c(\mu)L_A\log L_A+o(L_A\log L_A), \nonumber\\
&c(\mu)=\dfrac{\log2}{\pi^{2}(2\pi)^{d-1}}\int_{\partial\Omega}dS_{x}\int_{\partial\Gamma}dS_{k}|\mathbf{n}_{k}\cdot\mathbf{n}_{x}|,
 \end{align}
where $\mu$ is the chemical potential, $\Omega$ is the real space region $A$ and $\partial\Gamma$ is the Fermi surface. $\mathbf{n}_{x}$ and $\mathbf{n}_{k}$ denote the normal vectors on the spatial boundary $\partial\Omega$ and the Fermi surface respectively. For Eq.~\eqref{eq:klich}, the linear dimension of the system is scaled to unity. In this work, we will consider 37 fermions in 2D occupying momenta shown in Fig.~\ref{fig:fermisurface} for both free fermions and for a $\nu=1/2$ composite fermion non-Fermi liquid. 
A straight forward evaluation of $c(\mu)$ for the Fermi surface shown in Fig.~\ref{fig:fermisurface} and a square-shaped region $A$  results in 
an asymptotic form for the second R\'enyi entropy
\begin{equation}
S_{2, {\rm Widom}}\sim(0.134)\lambda \log \lambda
\label{eq:widom}
\end{equation}
 as a prediction based on ``Widom formula'' Eq.~\eqref{eq:klich}. From here on we use the dimensionless quantity $\lambda\equiv k_F L_A$, where $k_F$ is the radius of the Fermi surface.

{\it Monte Carlo evaluation of $S_2$.--} In order to calculate the R\'enyi entropy $S_2$ for 
the $\nu=1/2$ CF wave-function, we use 
the scheme of Ref.~\cite{melko2010} and consider two copies of the system 
to evaluate 
the expectation value of the SWAP operator 
which is related to $S_2$ 
as follows:
 \begin{align}
&e^{-S_{2}}\nonumber\\
&=\sum_{\beta_{1},\beta_{2}}\sum_{\alpha_{1},\alpha_{2}}\langle\beta_{2}|\langle \alpha_{1}|\Psi\!\rangle \!\langle \Psi|\alpha_{1}\rangle|\beta_{1}\rangle\langle\beta_{1}|\langle \alpha_{2}|\Psi\rangle\langle \Psi|\alpha_{2}\rangle|\beta_{2}\rangle\nonumber\\
&\equiv\left\langle {\rm SWAP}_A\right\rangle.
\end{align}
Here
$\alpha_i$ and $\beta_i$, with $i=1,2$ for the two copies, are real space coordinates within each copy of subregions, i.e., $\alpha_1\in A_1,\beta_1\in B_1$ and   $\alpha_2\in A_2,\beta_2\in B_2$. 
Below we calculate the expectation value for 
the model wave-function
by  sampling the  wave-function over the two copies,
introducing a ``particle number trick''
which improves the computing time and allows for parallelization compared to the previous calculation of $S_2$ for itinerant fermions\cite{mcminis} .

{\it Particle number trick.--} Compared to the case of positive definite spin wave-functions studied in Ref.~\cite{melko2010}, 
 itinerant fermion systems come with two major challenges against evaluation of $\langle {\rm SWAP}_A\rangle$: (1) the wave-function is not positive definite, (2) the number of fermions in the region $A$ fluctuates. The first issue had been partially mitigated in Ref.~\cite{vishwanath} using the so-called ``sign trick'' exactly factorizing $\langle {\rm SWAP}_A\rangle$  into a product of two terms each concerning only magnitude or only sign. On the other hand, the fermion number fluctuation was not an issue in Ref.~\cite{vishwanath} as the Gutzwiller projector ensured one fermion per site. \textcite{mcminis} dealt with the fermion number fluctuation for free fermions and for Slater-Jastrow trial wave-functions by discarding the Monte Carlo moves that result in different particle numbers for the region $A$ in the two copies. In this approach, the running time for free fermions scales as 
 $O(N^3)$ as a function of the number of particles $N$ when combined with 
 fast single-rank updates of the Slater determinant. However, we found a  
 direct adaptation of the algorithm used in Ref.~\cite{mcminis} for the CF wave-function to be prohibitively slow. 
This is mainly due to the fact that a change to the position of a single composite fermion changes all the elements of the determinant and hence we cannot benefit from the fast single-rank updates of the Slater determinant. This motivated us to develop an alternate way of dealing with the particle number fluctuation issue. 

We show in \cite{suppl} that $\langle {\rm SWAP}_A\rangle$
can be further exactly factorized into contributions from sectors of fixed particle numbers in each subregion. Evaluating the contributions from each sector separately not only effectively reduces the 
size of the space that needs to be sampled for each contribution but it also allows parallelization. Our algorithm takes about $10^4$ CPU hours per data point, compared to $10^5$ CPU hours per data point using the algorithm of Ref.~\cite{mcminis}. In addition the restriction of the sampled space achieved using the particle number trick reduces the error bars in a way similar to the "ratio trick" for lattice models proposed by \textcite{melko2010}.
Implementing the particle trick algorithm on a cluster of ~100 processors, we were able to obtain data for 20 different values of $l\equiv L_A/L$ with 37 composite fermions in days instead of weeks.

{\it Wave-function.--} 
The wave-function for the spin polarized $\nu=1/2$ can be written as a determinant of fermions in
zero field times a $\nu=1/2$ bosonic Laughlin state
\cite{PhysRevLett.72.900, PhysRevLett.84.4685}, which we write as 
\begin{equation}
\det_{i,j}t_i({\bf d}_j)|\Psi_L^{1/2}\rangle, 
\end{equation} where $t$ is a single electron translation operator, and ${\bf d}$ is a displacement
satisfying $N_\Phi {\bf d}\in m{\bf L}_1+n{\bf L}_2$, $m$ and $n$ are integers, and 
$N_\Phi$ is the magnetic flux quanta. The unit cell of the torus is specified by 
${\bf L}_1$ and ${\bf L}_2$, which spans an area equal to $2\pi N_\Phi \ell^2_B$, where $\ell_B$ is
the magnetic length. The
displacements are given in terms of wave-vectors of composite fermions by
$ k_a=\epsilon_{ab}d^b/\ell_B^2$. Acting with the determinant, the 
holomorphic part of the coherent state wave-function can be written as:
\begin{eqnarray}
{\big (}\sum_{P=1}^{N!} (-1)^P F_P(z_1,\ldots,z_N){\big )} F_{CM}(\sum_i(z_i-\bar{d})) \nonumber\\
F_P(\{z_i\})=\prod_{i<j}\sigma(z_i-z_j+d_{P(i)}-d_{P(j)})^2 \prod_ie^{d^*_{P(i)}z_i}
\label{eq:wf}
\end{eqnarray}
where $F_{CM}(z)=\sigma(z)^2$ is the center of mass wave-function, $d=(d_x+id_y)/\sqrt{2}\ell_B$, and $z=(x+iy)/\sqrt{2}\ell_B$  are complex distances and
coordinates. Note that in the definition of complex quantities, which are dimensionless, 
we include
a $\sqrt{2}$ factor. The function $\sigma(z)$ is  the Weierstrass $\sigma$ function which
in terms of the Jacobi $\vartheta$ function is:
\begin{equation}
\sigma(z)={\vartheta_1(\kappa z;\tau)\over \kappa \vartheta_1^\prime(0;\tau) }
\exp(i(\kappa z)^2/\pi(\tau-\tau^*)).
\end{equation} 
Here $\kappa=\pi/L_1$, $L=(L_x+L_y)/\sqrt{2}\ell_B$ is the linear dimension of the system
with $L^*_1L_2-L^*_2L_1=2\pi i N_\Phi$, and $\tau=L_2/L_1$ is the modular parameter of the torus.

The wave-function in Eq.~\eqref{eq:wf}
cannot be cast as a determinant and therefore proves 
inconvenient for Monte-Carlo calculations.  Instead, we present a new expression that is inspired
by a construction due to Jain and Kamilla\cite{PhysRevLett.63.199, PhysRevB.55.R4895}.  These authors 
treat the translation operators as $c$-numbers and take the Jastrow factors  
inside the determinant.  Doing so reduces their action on 
$\sigma(z_i-z_k)$ (the relative part) from $2(N-1)$ to $N-1$. 
To compensate we double each $d_i(t_j)$.

\begin{eqnarray}
F_{CF}=\det_{i,j} (e^{d^*_j z_i} \prod_{k(\neq i)}\sigma(z_i-z_k+2(d_j-\bar{d}))\times \nonumber\\
F_{CM}(\sum_i(z_i-\bar{d})).
\label{eq:CFwf}
\end{eqnarray} 
To obtain the full  expression of the  wave-functions a non-holomorphic exponential factor
$e^{-\sum_i z_iz^*_i/2}$ has to be included. For convenience we have made a specific
choice of the zeros of the center of mass part of the 
wave-function\cite{PhysRevB.31.2529}. This resolves the two-fold topological degeneracy 
of the state.

For the $\nu=1/2$ state in a  magnetic field it is non-trivial to decide how to fill the ``Fermi sea'', as the usual kinetic energy is completely quenched due to the magnetic field. 
One of us has proposed a model Hamilton\cite{haldaneunpub}, which can be used to find the composite fermion Fermi sea.
\begin{equation}
H_{\rm ``kin''} \equiv\frac{\hbar^2}{2mN}\sum_{r<s}|{\bf k}_r-{\bf k}_s|^2
\end{equation}
where $m$ is the electron mass.
The above ``kinetic energy'' is independent of uniform boost (the so called k-invariance) as is the energy of the model wave-function. The invariance under uniform boost guarantees $F_1=-1$\cite{PhysRevLett.79.4437,PhysRevLett.80.5457,PhysRevB.58.16262}, which is the only nonzero Fermi liquid parameter of the model. Hence by choosing the momenta ${\bf k}_r$'s that minimize this ``kinetic energy'', the wave-function effectively describes a state with diverging effective mass as indicated by the non-perturbative effect of electron coupling to the fluctuating Chern-Simons gauge field in field theoretic studies\cite{PhysRevB.47.7312,Nayak1994359}. For particular values of $N$, including $N=37$, we consider the set of momenta minimizing $H_{\rm ``kin''} $ to be same as the set of momenta filling the Fermi sea for free fermions minimizing the usual kinetic energy. Hence we can use the same set of momenta shown in Fig.~\ref{fig:fermisurface} for both the composite and free fermion wave-functions. The pair correlation function calculated using the composite fermion wavefunction Eq.~\eqref{eq:CFwf} with this choice of momenta show Fermi-Liquid-like $2k_F$ oscillations~\cite{PhysRevLett.72.900}.

\begin{figure}[t]\centering
         \includegraphics[width=.45\textwidth]{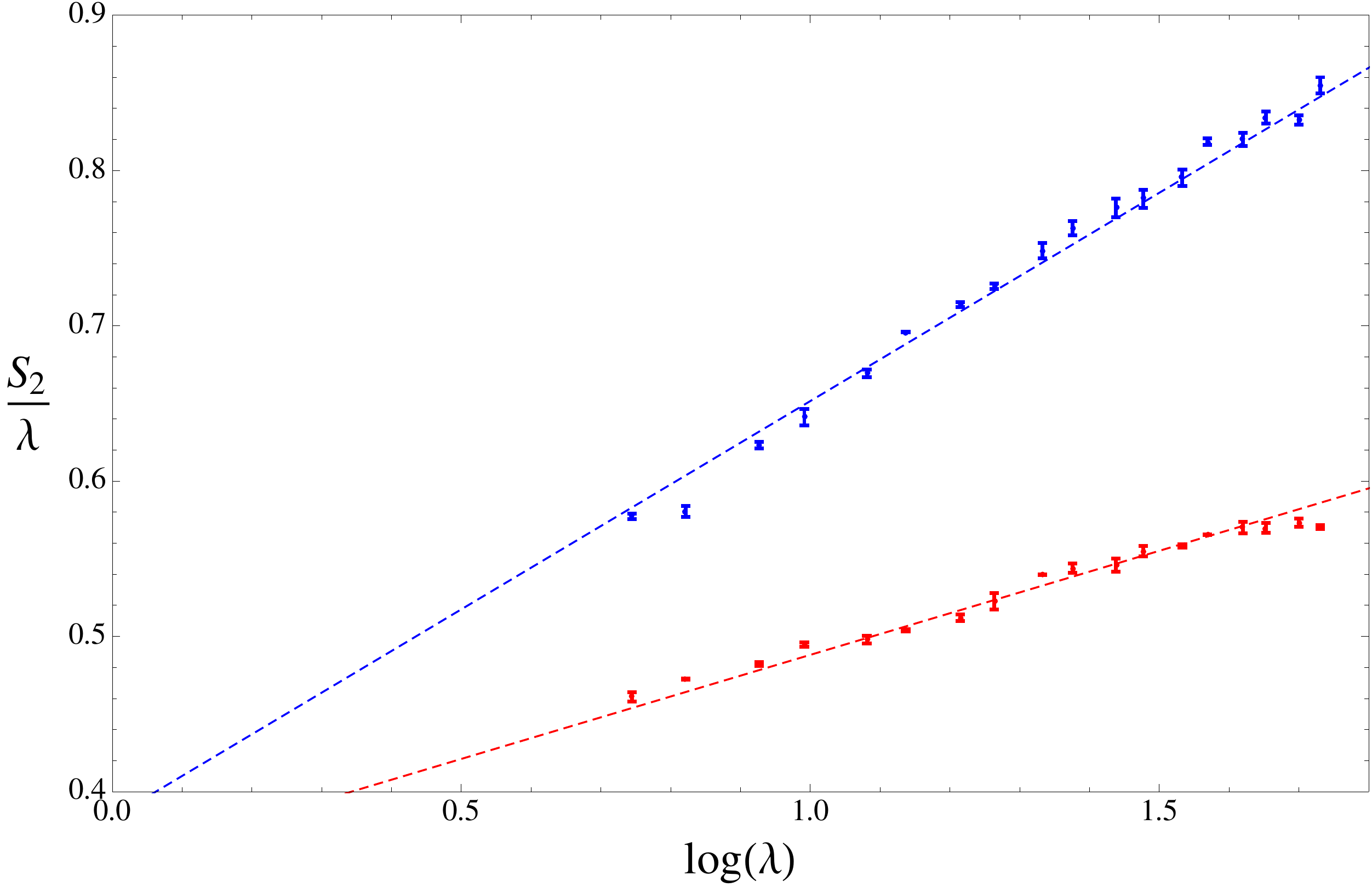}
         \caption{Plot of $S_2/\lambda$ as a function of $\lambda$.  Red corresponds to free fermions, while blue corresponds to composite fermions at half-filling in the lowest Landau level. Error bars indicate 95\% confidence intervals. Dashed lines are best linear fits of $S_2/\lambda$ to the form $a+c \log \lambda$, where $a$ and $c$ are fit parameters.
                        }
        \label{fig:results}
      \end{figure}

{\it Results.-} - We obtained the second R\'enyi entropy $S_2$
as a function of  $\lambda$  for both the free fermion gas and the composite fermion trial wave-function for $\nu=1/2$ state. Here $L_A$ is the linear dimension of the system and $k_F=\sqrt{10\pi/37}$ is the radius of the Fermi surface. 
The results and the error bar are shown in Fig.~\ref{fig:results}. 
Our results for the free fermion gas shown in red are in good agreement with previous numerical results\cite{calabrese,mcminis} and hence demonstrate reliability of our algorithm using the particle number trick. 
 The ratio $S_2/\lambda$ is linear in $\log \lambda$ and the linear fit results in 95\% confidence level shown in the red dashed line in Fig.~\ref{fig:results} follows 
\begin{equation}
S_2[\Psi_0] =(0.135\pm 0.01) \lambda\log \lambda
\label{eq:free}
\end{equation}
where $\Psi_0$ denotes the free Fermi gas wave-function for the set of momenta shown in Fig.~\ref{fig:fermisurface}. 
Comparing the results Eq.~\eqref{eq:free} to
 Eq.~\eqref{eq:widom}
we again confirm the validity of the Widom formula Eq.~\eqref{eq:klich} for free fermions.

For the $\nu=1/2$ non-Fermi liquid state, the results shown in blue in Fig.~\ref{fig:results} again exhibit linear dependence of $S_2/{\lambda}$ on $\log{\lambda}$, i.e. multiplicative logarithmic violation of the area law. However, the linear fit again at 95\% confidence level: 
\begin{equation}
S_2[\Psi_{\nu=1/2}]= {(0.27\pm0.02) \lambda \log \lambda}
\label{eq:CF}
\end{equation}
reveals that the coefficient is no longer given by the ``Widom formula''. The steeper slope for the non-Fermi liquid state is evident even from the raw data. Curiously, the coefficient of the multiplicative logarithmic correction term for  $\nu=1/2$ non-Fermi liquid is not only larger compared to that for free fermions, but it appears to be double the value expected from the ``Widom formula'' Eq.~\eqref{eq:klich}.

{\it Discussion.--} In this letter we introduced a particle number trick which can speed up variational Monte Carlo calculation of R\'enyi entropies via parallelization and used the improved algorithm to calculate the second R\'enyi entropy $S_2$ of the $\nu=1/2$ composite fermion non-Fermi liquid state captured by the trial wave-function. We found the multiplicative logarithmic violation of the area law with the same functional dependence on the linear dimension of the subregion, i.e., $S_2\propto {\lambda\log\lambda}$, as is the case for free fermions
 but with a coefficient that is roughly double what is found for free fermions. Our results support the conjecture that $S_2\propto {\lambda\log\lambda}$ might be the strongest form of area law violation in 2D made in Ref.~\cite{swinglesenthil}. 
However against the conjecture made in \cite{PhysRevB.86.035116} regarding universality of the Widom formula,
the strong enhancement of the entanglement entropy captured by coefficient doubling in our results revealed a violation of  the Widom formula for the $\nu=1/2$ non-Fermi liquid state. 

Our explicit calculation of entanglement entropy for an established non-Fermi liquid state raises a plethora of interesting questions.  One question is how generic are such factors of order one change in the coefficient of $L_A\log L_A$ term in the entanglement entropy of strongly correlated fermions forming a non-Fermi liquid state. There is little literature on this coefficient for interacting fermions. While a Gutzwiller projected Fermi surface constrained to maintain one fermion per site showed little difference from free fermions\cite{vishwanath}, Slater-Jastrow wave-functions showed small changes in the coefficient\cite{mcminis}. However, while  Slater-Jastrow wave-functions are frequently used as a way of building in correlation effects to wave-functions and used to model Fermi liquids, it is not clear if the effect of the Jastrow factor is perturbative as the $S_2$ result of Ref.~\cite{mcminis} does not extrapolate to free fermion result in the limit of fermion residue $Z\rightarrow 1$. It will be interesting to use the particle number trick on a d-wave metal wave-function which is proposed to be stabilized by a ring-exchange Hamiltonian\cite{Jiang:2013fk} in this context. Another question is whether it is possible to gain analytic insight into the enhancement of entanglement due to strong correlation we found, building on the field theory literature on the $\nu=1/2$ state. Finally, the investigation of entanglement spectra may reveal more dramatic differences between the free fermions and $\nu=1/2$ non-Fermi liquid state.

\noindent {\bf Acknowledgements}
We thank Tarun Grover, Roger Melko, Nick Read and Abolhassan Vaezi, for helpful discussions. F.D.M.H. and E.H.R. were supported by DOE Grant DE-SC0002140,  E.-A.K. was supported by NSF CAREER grant DMR 0955822, J.S. was supported by NSF-DMR 0955822 during his internship at Cornell University. 
This work used the Extreme Science and Engineering Discovery Environment (XSEDE), 
which is supported by National Science Foundation grant number OCI-1053575 and 
Computational Center for Nanotechnology Innovations (CCNI) through NYSTAR. 
F.D.M.H. also acknowledges support from the W. M. Keck Foundation. This work was also partially supported by a grant from the Simons Foundation
(\#267510 to F. D. M. Haldane)  for Sabbatical Leave support.

\bibliography{ref}

\end{document}